Thermoelectric Power of Bi and Bi-Te(0.14%) in Porous Vycor Glass

T. Huber

Laser Laboratory, Howard University, Washington, DC,

A. Nikolaeva, A. Gitsu, D. Konopko

Institute of Solid State Physics, Academy of Sciences, Moldova,

and

M. J. Graf and J. Huang,

Department of Physics, Boston College, Chestnut Hill, MA 02467

Semiconductor quantum wires constitute a promising thermoelectric material because of the increase of the electronic density of states in low-dimensional materials. We studied the magnetic-field-dependent resistance and Seebeck coefficient of a high-density network of 6-nm-diameter wires of Bi and of Bi doped with 0.14% component of Te embedded in porous Vycor glass. The resistance R increases as temperature decreases from 300 K down to 0.3 K for both composites. However, in contrast to recent results that demonstrate the semiconducting behavior of the resistance and very large enhancements of the thermoelectric power of composites containing Bi nanowires with diameters of 9 and 15 nm, we find that the resistance is not thermally activated and that the composites' thermoelectric power is of the same order of magnitude as the thermoelectric power of bulk Bi. Our results are consistent with the nanowires having carrier density that is enhanced by surface effects.

## SECTION I. INTRODUCTION

Progress in the synthesis of nanostructured materials is advancing at a rapid pace. One of the most exciting prospects is that of an ideal quantum wire of diameter $d$ that is less than the Fermi wavelength $\lambda_F$. Profound changes in the density of state (DOS) and deep modifications of the transport properties are expected for such nanowires. In this vein, the study of the transport properties of carbon nanotubes has been pursued since their discovery in 1991—with considerable success.[1] Comparatively little attention has been given to the semimetal Bi, a material with near-ideal properties for thermoelectric applications at ~100 K. Bi has a very small effective mass, resulting in a long Fermi wavelength (~25 nm), which eases the restrictions on the fabrication of quantum wires. Bulk Bi is a semimetal whose three conduction band minima at the L-points overlap the valence band maximum at the T point by about 40 meV. The temperature coefficient of the resistance of crystalline Bi is positive, i.e., the element is metallic. The thermoelectric power, which is anisotropic, is roughly between -50 and -100 µV/K at 100 K (the negative sign indicates that the electron mobility is larger that the hole mobility).[2,3] Recently, Heremans *et al.* studied the transport properties and the Seebeck coefficient of Bi nanowire composites that were synthesized by the vacuum impregnation of Bi in porous alumina and silica templates with pore diameters between 9 and 15 nm.[4] They report that the composites have a very high resistivity of ~500 MΩ.cm at 100 K, a negative temperature coefficient of the resistance, and very large thermoelectric power (at 100 K, $|S| \sim 10^4$ and $10^5$ µV/K for 15- and 9-nm samples, respectively). Presumably, their results agree with the confinement theory by Dresselhaus *et al.,*[5,6] which is a careful evaluation of the density-of-states (DOS) of electrons and holes in Bi nanowires. As a result of an increase in the zero-point energy of the electrons, the band overlap decreases for thin wires, and if the wire diameter is smaller than a critical diameter $d_c$ (where $d_c \sim \lambda_F$), a gap is formed and the material undergoes a semimetal-to-semiconductor transition (SMSC). This theory predicts that the maximum enhancement of the thermoelectric power increases with decreasing wire diameter, although the actual value of S depends upon the charge carrier density in the nanowires. Here, we present a study of a network of 6-nm Bi quantum wires in porous Vycor glass (PVG), a prototype monolithic nanoscale porous material that has been used extensively as a confinement matrix.[7] We find that the temperature coefficient of the resistance is negative. However, in contrast to the results described in reference 4, we find that the resistivity and thermoelectric power of the composites are of the same order of magnitude as those found for bulk Bi. The resistivity has been presented in previous publications.[8,9]

Sandomirski predicted the SMSC transition of thin films on the basis of a model of quantum confinement.[10] Despite many transport and optical investigations of quantum phenomena in thin films, the experimental evidence is not conclusive. A semiconductor is intrinsic if its electronic properties are dominated by electrons that are thermally excited from the valence band and extrinsic if its electronic properties are dominated by carriers contributed to the conduction band by impurities (or captured from the valence band by impurities). The resistivity of these films is weakly T dependent, unlike a typical intrinsic semiconductor that exhibits a thermally activated behavior [($\rho \sim e^{-\Delta/2T}$ ), where $\Delta$ is the energy gap].[11] This has been interpreted in terms of a very short effective mean free path $l \sim t$ as well as a surface-enhanced carrier density.[12,13] Here, $t$ is the film thickness. Since the majority carrier concentration then becomes insensitive to the introduction of an energy gap, one would expect no abrupt changes at the transition point. It is clear that Bi nanowires share many characteristics with Bi films, including the close contact with an insulating substrate and the quantum confinement. We will show that the dependence of resistivity on temperature is a property shared by very fine nanowires and films alike, and we will show that the model of surface charges can explain our experimental results.

The plan of the paper is as follows. In Sec. II, we briefly discuss the sample preparation method, sample characterization, and other experimental issues. In Sec. III, we analyze the experimental results and present our conclusions.

# SECTION II. SAMPLE PREPARATION

The network of Bi quantum wires is fabricated by the template injection technique.[14] The template used in this work is porous Vycor glass (PVG), a widely used prototype monolithic nanoscale porous material from Corning. The narrow pore size distribution of PVG has made it the material of choice for many studies of the effects of the porous environment on diffusion, reaction, and other physical and chemical properties. The average pore diameter of our PVG templates is 6 nm. The interconnected network of pores occupies approximately 30% of the total volume. While the silica backbone structure is complex and interconnected, one can consider it as being made up of silica particles with a characteristic size of 26 nm. The use of PVG as a confining host has been particularly fruitful, for example, in studying the properties of superfluid helium[7] and that of superconductors such as In[15] in a restricted geometry. The network structure of PVG, which is the result of a spinodal decomposition of the glass, is well known from small-angle X-ray scattering studies[16] and small-angle neutron scattering (SANS).[17]

In most cases, molten metals do not wet glass or alumina, and an external pressure is required to force impregnation of the insulating preform with a liquid metal. The pressure is $P = - 2\gamma_{lv} \cos \theta / R$, where R is the pore radius and $\theta$ is the contact angle between the liquid metal and the insulator surface. For $\theta > 90°$, the system is nonwetting. The surface tension of liquid Bi, $\gamma_{lv}$, is 380 dyn/cm. Assuming the least favorable case of nonwetting ($\theta = 90°$), Washburn's equation gives $P = 6/d$, where P is measured in kbar and $d$ is the pore diameter in nanometers. Thus, theoretically, a minimum applied pressure of 1 kbar is required to fill the PVG with molten Bi. We have observed that for Bi-PVG there is such a threshold pressure, $P_t$, and that if the injection is conducted for $P < P_t$, the amount of Bi in the composite decreases dramatically. The composites synthesized under such conditions can be distinguished from full Bi-PVG composites by measuring the density. Also, if the pressure is too low ($P << P_t$) the material, which is black when full, becomes less opaque. Here, we employed a pressure of 5 kbar for preparing the composites and estimated that all channels larger than 2.7 nm in diameter will be filled with liquid Bi. The samples were prepared from Bi of 99.999% purity.

Figure 1 shows a scanning electron microscope (SEM) image of the composite. The Bi in the composite was exposed by etching with HF. We have used this technique previously for In and for other materials.[15] The shiny black samples have approximately 80% of the pore volume filled with Bi. Figure 2 shows the transmission electron microscope image of the composite, which shows that the insulator phase consists of spherical particles of silica with an average diameter of 26 nm but that the shape and location of the particles are not regular. The SANS data of certain controlled-pore glasses, which are analogous to PVG in that they result from spinodal decomposition, have been modeled by using a stochastic superposition of plane waves,[16] and the results are shown in Fig. 3. X-ray diffraction (XRD) from the Bi-PVG composite shows that the Bi retains its rhombohedral (trigonal) structure but with shrinkage of its unit cell. For example, the (102) planes in the hexagonal indexing system, which are separated by 0.328 nm in bulk Bi, are separated by 0.320 nm in the composite. This corresponds to a lattice linear contraction of approximately 2.5%. This is likely a result of the injection process, since Bi expands by 3.3% on solidification. An estimate for the average Bi crystallite size, $D$, can be obtained from the widths of the XRD peaks by using Scherrer's equation.[18] A value of $D = 9$ nm was determined from the 0.017 full-width and half maximum (FWHM) of the peak corresponding to the (102) planes, comparable to the PVG average pore diameter of $d$=6 nm. We have observed that these samples are not very stable and that samples larger than a few millimeters crumble when subjected to stress. Still, we were able to select samples that were robust and therefore suitable for our experiments.

## III. EXPERIMENTAL RESULTS AND DISCUSSION

The sample resistance was measured in a gas-flow cryostat operating in a temperature range of 2 to 300 K using four terminal DC and AC (f = 100 Hz) techniques.

Typically, the samples were 3 mm in length and have a cross-section of 2 mm × 1 mm. Electrical contact was made via brass wires attached with silver epoxy to gold pads deposited in a vacuum evaporator. The composite resistivity is R*A/L*, where *A* and *L* are the appropriate cross-sectional surface and length. The temperature-dependent composite resistivity of the Bi-PVG and the Bi-Te-PVG samples is shown in Fig. 4. In comparison, the bulk Bi resistivity obeys a $T^2$ law at low temperatures and is roughly proportional to T for T > 100 K. At 300 K, the resistivity of the Bi composite is roughly 21 mΩ-cm, and cooling to 4 K increases this to 27 mΩ-cm. The ratio of the room-temperature resistivity of the PVG-Bi composite to that of the bulk is $r = \rho_{Bi\text{-}PVG}/\rho_{bulk} = 175$. Taking into account that the material in the composite occupies 80% of the pore volume and that about 1/3 of the metal wires are oriented along the current flow, we find that the resistivity of the material in the Vycor is actually larger by a factor of 15 than that of the composite. Such a geometric correction works fairly well for an In-filled 6-nm PVG template, where the electronic mean free path $l_e << d$ and surface scattering is not very important,[8] and *r* is experimentally determined to be about 20 at room temperature. Therefore, the resistivity of the Bi composite is roughly one order of magnitude larger than expected from structural considerations. The temperature dependences of Bi-PVG and Bi-Te-PVG have been discussed in a previous publication.[15] The electronic transport results can be understood in terms of a surface-enhanced carrier density. In thin films,[12,13] it was found that the surface enhanced carrier density, $3 \times 10^{12}$ holes/cm$^2$, is the effective density per unit film area. Considering that the number of surface states is proportional to the surface area, the effective hole bulk density can be estimated to be $5 \times 10^{18}$ holes/cm$^3$ for 6-nm wires. Assuming that the transport properties are controlled by a single carrier, we find that $\rho_{Bi\text{-}PVG} = \dfrac{1}{nq\mu}$ . We obtain $\mu = 10^3$ cm$^3$ /(sec volt), a value that is in good agreement with thin Bi film results of $t = 20$ nm. This effect would not explain the discrepancy between our results and those in reference 4 because the effect should be roughly the same for both types of nanowire samples.

The expected semiconductor activation energy of 6-nm Bi nanowires can be obtained as follows. The Bi nanowire's gap has been calculated in reference 6. Results are presented for wire diameter $d$ between 16 and 200 nm. The gap is found to obey $\Delta = 37\,meV(d_c / d\ -1)$, where $d_c$ = 40 nm for wires along the trigonal axis and 60 nm for wires along the bisectrix axis. We find that the gap for our 6-nm wires is predicted to range between 0.2 and 0.3 eV. Another estimate can be obtained as follows. The presumed thin film's gap dependence was experimentally[12] found to obey roughly $\Delta = 40$ meV $((t_c/t)^2 -1)$, where $t$ is the film thickness and $t_c$= 30 nm. Assuming the behavior of a nanowire of $d = 6$ nm to be that of a thin film of the same thickness, we obtain $\Delta = 1$ eV. As shown in Fig. 4, the Bi wire network fails to display the thermally activated behavior expected for an intrinsic semiconductor with band gaps in the range between 0.2 and 1 eV because such a fit, applied to the Bi-PVG data, is only reasonable over a highly restricted temperature interval.

The samples' thermopower was measured in a closed-cycle refrigerator operating in a temperature range of 3 to 300 K. The differential thermopower between the samples and copper is defined as $S = V/(T_H - T_C)$, where $T_H - T_C$ is the temperature difference established and $V$ is the potential difference generated between the ends of the sample. We employed the arrangement shown in the inset of Fig. 5. The copper blocks are 1 mm × 1 mm. The composite samples are approximately 1 mm in length and the bulk Bi samples are 5 mm in length. A Cu-CuFe$_{0.01at.\%}$ thermocouple is in thermal contact but electrically insulated from each copper block. The approach employed has several sources of error. Heat may be conducted along the wires to the sample, thereby changing the temperature in the areas of contact. To eliminate this error, the thermocouple's wires are thin (diameter = 60 μm) and the leads are thermally anchored to the copper blocks. Another type of error occurs if the contact thermal resistance between the copper blocks and sample is higher than the sample thermal conductivity. This error is minimized by employing a low-melting-temperature In$_{0.5}$Ga$_{0.5}$ eutectic solder. The contact thermal resistance across the copper-sample interface is estimated to be $10^2$ K/W at 4 K.[19] The sample thermal conductivity, which has not been measured, can be estimated as follows. The thermal conductivity κ of polycrystalline Bi is around 3 W•cm$^{-1}$•K$^{-1}$.[20] The sample thermal conductivity is less than that of polycrystalline Bi on account of the fractional filling factor and of the much-reduced mean free path of the carriers in the composite in comparison to the bulk. The thermal resistance of the PVG template, which is estimated to be $10^3$ K/W at 4 K, is much higher than the thermal resistance of the Bi nanowire network and can therefore be neglected. $\Delta T$ is fixed at approximately 3 K at all temperatures.

The thermopower of the Bi and Bi-Te composites is shown in Fig. 5 along with the thermopower of a single-crystal Bi sample for two orientations of the C3 axis with respect to the direction of heat flow. The measured single-crystal thermopower is consistent with previous measurements, considering that $\Delta T \sim 3$ K. The composites' thermopower values are roughly 50 and 15% of the thermoelectric power of the single-crystal sample when oriented with the C3 axis perpendicular and parallel to the heat flow direction, respectively. This is in sharp contrast to the findings of reference 4, in which a several-order-of-magnitude enhancement of the thermoelectric power of the 9- and 15-nm composites was noted. The negative sign of the Seebeck coefficient can be understood as indicating that the majority carriers are electrons. This is surprising because, in thin films of Bi on borosilicate glass, it was found that the majority carriers are holes. However, Heremans *et al.* observed that the sign of the Seebeck coefficient of their composites is negative for silica-embedded Bi nanowires and positive for alumina-embedded Bi nanowires—and it is reasonable to expect that the sign of the charge of the impurities states depends upon the details of the interaction between the Bi and the surface electric fields. The absolute value of the thermoelectric power of Bi-Te composites is larger than the thermoelectric power of Bi composites, which is consistent with significant two-carrier conduction in pure Bi samples considering the smaller electrical conductivity of Bi-Te samples in comparison with Bi samples and the increase in the carrier density for Bi-Te composites in comparison with Bi composites.

It is obvious that Bi is subject to large stresses in the pores of PVG and that these might induce a phase change. A hydrostatic pressure of 20 kBar induces the transition from semimetal Bi(I) to the metallic Bi(II). We can calculate the stresses present in our samples in the following manner: from a linear contraction of 2.5% in the (102) direction and taking Young's modulus to be $5 \times 10^{11}$ dyn/cm,[2,19] we find a stress of $1.3 \times 10^9$ dyn/cm$^2$ or 13 kBar. This pressure is insufficient to promote the Bi(I)-(II) transition.[20] Also, the effect of compression in thick (500-nm) Bi films has been investigated, and it is known that a pressure of 10 kBar increases the film resistivity by only 7%.[21] Therefore, we believe that the stresses in our samples do not significantly affect our results.

The sample resistance was measured in a gas-flow cryostat operating in a temperature range of 2 to 300 K using four terminal DC and AC ($f = 100$ Hz) techniques. Typically, the samples were 3 mm in length and have a cross-section of 2 mm × 1 mm. Electrical contact was made via brass wires attached with silver epoxy to gold pads deposited in a vacuum evaporator. The composite resistivity is R$A/L$, where $A$ and $L$ are the appropriate cross-sectional surface and length. The temperature-dependent composite resistivity of the Bi-PVG and the Bi-Te-PVG samples is shown in Fig. 4. In comparison, the bulk Bi resistivity obeys a $T^2$ law at low temperatures and is roughly proportional to T for T > 100 K. At 300 K, the resistivity of the Bi composite is roughly 21 mΩ-cm, and cooling to 4 K increases this to 27 mΩ-cm. The ratio of the room-temperature resistivity of the PVG-Bi composite to that of the bulk is $r = \rho_{\text{Bi-PVG}}/\rho_{\text{bulk}} = 175$. Taking into account that the material in the composite occupies 80% of the pore volume and that about 1/3 of the metal wires are oriented along the current flow, we find that the resistivity of the material in the Vycor is actually larger by a factor of 15 than that of the composite. Such a geometric correction works fairly well for an In-filled 6-nm PVG template, where the electronic mean free path $l_e \ll d$ and surface scattering is not very important,[8] and $r$ is experimentally determined to be about 20 at room temperature. Therefore, the resistivity of the Bi composite is roughly one order of magnitude larger than expected from structural considerations. The temperature dependences of Bi-PVG and Bi-Te-PVG have been discussed in a previous publication.[15] The electronic transport results can be understood in terms of a surface-enhanced carrier density. In thin films,[12,13] it was found that the surface enhanced carrier density, $3 \times 10^{12}$ holes/cm$^2$, is the effective density per unit film area. Considering that the number of surface states is proportional to the surface area, the effective hole bulk density can be estimated to be $5 \times 10^{18}$ holes/cm$^3$ for 6-nm wires. Assuming that the transport properties are controlled by a single carrier, we find that $\rho_{\text{Bi-PVG}} = \frac{1}{nq\mu}$ . We obtain $\mu = 10^3$ cm$^3$ /(sec volt), a value that is in good agreement with thin Bi film results of $t = 20$ nm. This effect would not explain the discrepancy between our results and those in reference 4 because the effect should be roughly the same for both types of nanowire samples.

The expected semiconductor activation energy of 6-nm Bi nanowires can be obtained as follows. The Bi nanowire's gap has been calculated in reference 6. Results are presented for wire diameter $d$ between 16 and 200 nm. The gap is found to obey $\Delta = 37 meV(d_c / d - 1)$, where $d_c.= 40$ nm for wires along the trigonal axis and 60 nm for wires along the bisectrix axis. We find that the gap for our 6-nm wires is predicted to range between 0.2 and 0.3 eV. Another estimate can be obtained as follows. The presumed thin film's gap dependence was experimentally[12] found to obey roughly $\Delta = 40$ meV $((t_c/t)^2$-1), where $t$ is the film thickness and $t_c.= 30$ nm. Assuming the behavior of a nanowire of $d = 6$ nm to be that of a thin film of the same thickness, we obtain $\Delta = 1$ eV. As shown in Fig. 4, the Bi wire network fails to display the thermally activated behavior expected for an intrinsic semiconductor with band gaps in the range between 0.2 and 1 eV because such a fit, applied to the Bi-PVG data, is only reasonable over a highly restricted temperature interval.

The samples' thermopower was measured in a closed-cycle refrigerator operating in a temperature range of 3 to 300 K. The differential thermopower between the samples and copper is defined as S = V/(T$_H$ − T$_C$), where T$_H$ − T$_C$ is the temperature difference established and V is the potential difference generated between the ends of the sample. We employed the arrangement shown in the inset of Fig. 5. The copper blocks are 1 mm × 1 mm. The composite samples are approximately 1 mm in length and the bulk Bi samples are 5 mm in length. A Cu-CuFe$_{0.01at.\%}$ thermocouple is in thermal contact but electrically insulated from each copper block. The approach employed has several sources of error. Heat may be conducted along the wires to the sample, thereby changing the temperature in the areas of contact. To eliminate this error, the thermocouple's wires are thin (diameter = 60 μm) and the leads are thermally anchored to the copper blocks. Another type of error occurs if the contact thermal resistance between the copper blocks and sample is higher than the sample thermal conductivity. This error is minimized by employing a low-melting-temperature $In_{0.5}Ga_{0.5}$ eutectic solder. The contact thermal resistance across the copper-sample interface is estimated to be $10^2$ K/W at 4 K.[19] The sample thermal conductivity, which has not been measured, can be estimated as follows. The thermal conductivity κ of polycrystalline Bi is around 3 W•cm$^{-1}$•K$^{-1}$.[20] The sample thermal conductivity is less than that of polycrystalline Bi on account of the fractional filling factor and of the much-reduced mean free path of the carriers in the composite in comparison to the bulk. The thermal resistance of the PVG template, which is estimated to be $10^3$ K/W at 4 K, is much higher than the thermal resistance of the Bi nanowire network and can therefore be neglected. ΔT is fixed at approximately 3 K at all temperatures.

The thermopower of the Bi and Bi-Te composites is shown in Fig. 5 along with the thermopower of a single-crystal Bi sample for two orientations of the C3 axis with respect to the direction of heat flow. The measured single-crystal thermopower is consistent with previous measurements, considering that ΔT ~ 3 K. The composites' thermopower values are roughly 50 and 15% of the thermoelectric power of the single-crystal sample when oriented with the C3 axis perpendicular and parallel to the heat flow direction, respectively. This is in sharp contrast to the findings of reference 4, in which a several-order-of-magnitude enhancement of the thermoelectric power of the 9- and 15-nm composites was noted. The negative sign of the Seebeck coefficient can be understood as indicating that the majority carriers are electrons. This is surprising because, in thin films of Bi on borosilicate glass, it was found that the majority carriers are holes. However, Heremans *et al.* observed that the sign of the Seebeck coefficient of their composites is negative for silica-embedded Bi nanowires and positive for alumina-embedded Bi nanowires—and it is reasonable to expect that the sign of the charge of the impurities states depends upon the details of the interaction between the Bi and the surface electric fields. The absolute value of the thermoelectric power of Bi-Te composites is larger than the thermoelectric power of Bi composites, which is consistent with significant two-carrier conduction in pure Bi samples considering the smaller electrical conductivity of Bi-Te samples in comparison with Bi samples and the increase in the carrier density for Bi-Te composites in comparison with Bi composites.

It is obvious that Bi is subject to large stresses in the pores of PVG and that these might induce a phase change. A hydrostatic pressure of 20 kBar induces the transition from semimetal Bi(I) to the metallic Bi(II). We can calculate the stresses present in our samples in the following manner: from a linear contraction of 2.5% in the (102) direction and taking Young's modulus to be $5 \times 10^{11}$ dyn/cm,[2,19] we find a stress of $1.3 \times 10^9$ dyn/cm$^2$ or 13 kBar. This pressure is insufficient to promote the Bi(I)-(II) transition.[20] Also, the effect of compression in thick (500-nm) Bi films has been investigated, and it is known that a pressure of 10 kBar increases the film resistivity by only 7%.[21] Therefore, we believe that the stresses in our samples do not significantly affect our results.

## IV. SUMMARY

In summary, the results of an experimental study of the resistance and the thermoelectric power of a network of 6-nm Bi and Bi-Te wires are reported. The nanowires are embedded in porous Vycor glass, a well-known porous material. The results are in sharp contrast to previous results, which reported a large increase of the thermoelectric power consistent with theoretical predictions of the Dresselhaus model of confinement. We found that the temperature dependence of the thermoelectric power of the nanocomposites is metallic ($dS/dT > 0$) and can be understood in terms of a model of excess surface charges that circumvent the SMSC transition.

FIGURE CAPTIONS

Fig. 1.  Scanning electron photomicrograph showing the surface of the 6-nm Bi nanowire composite. Dark areas are Bi.

Fig. 2.  Transmission electron microscope (TEM) photomicrograph of a grain of the Bi-PVG composite showing the silica particles. The dark areas are Bi. The average Bi particle size is 10 nm. The image height is 120 nm.

Fig. 3.  The network structure of porous Vycor glass reported in the literature. The figure is from reference 17 and corresponds to CPG 70 that has a filling factor similar to PVG.

Fig. 4.  Temperature dependence of the resistance of the samples of Bi-silica (solid line) and Bi-Te-silica (dashed line). The dotted lines represent fits to the Bi-PVG data of the form $R = R_o \exp(-\Delta/2T)$, with energy gap values $\Delta$ as indicated.

Fig. 5.  Thermoelectric power of the Bi-PVG sample (full circles) and Bi-Te-PVG sample (open circles). The thermopowers of single-crystal Bi with $\Delta T//C_3$ (solid line) and $\Delta T \perp C_3$ from Ref. 2 (dashed line) are also shown.

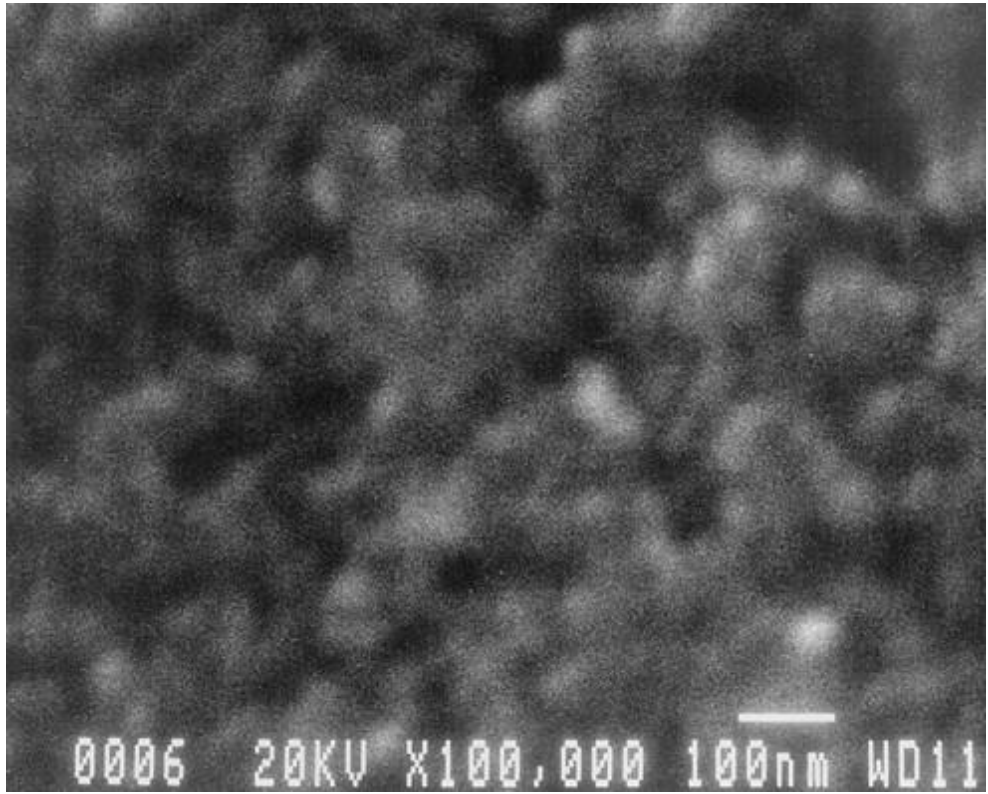

Thermoelectric Power of Bi and Bi-Te(0.14%) in Porous Vycor Glass  Huber *et al* .

Fig. 1

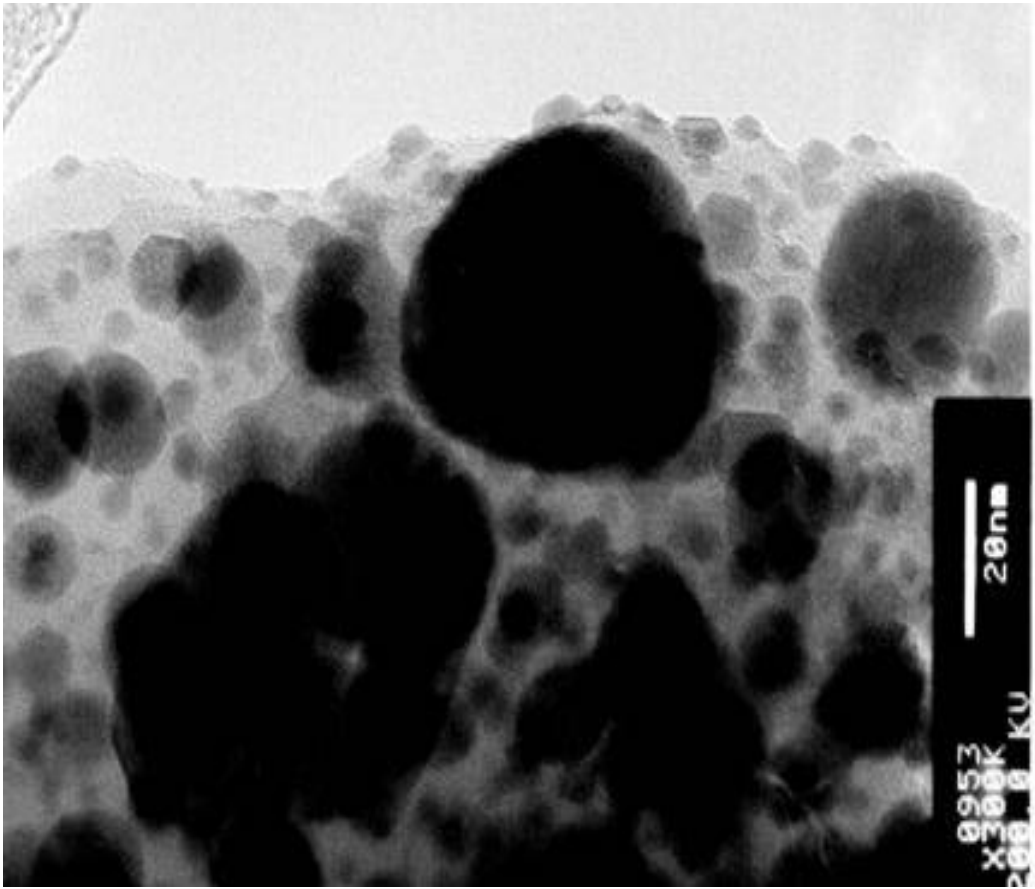

Thermoelectric Power of Bi and Bi-Te(0.14%) in Porous Vycor Glass  Huber *et al* .
Fig. 2

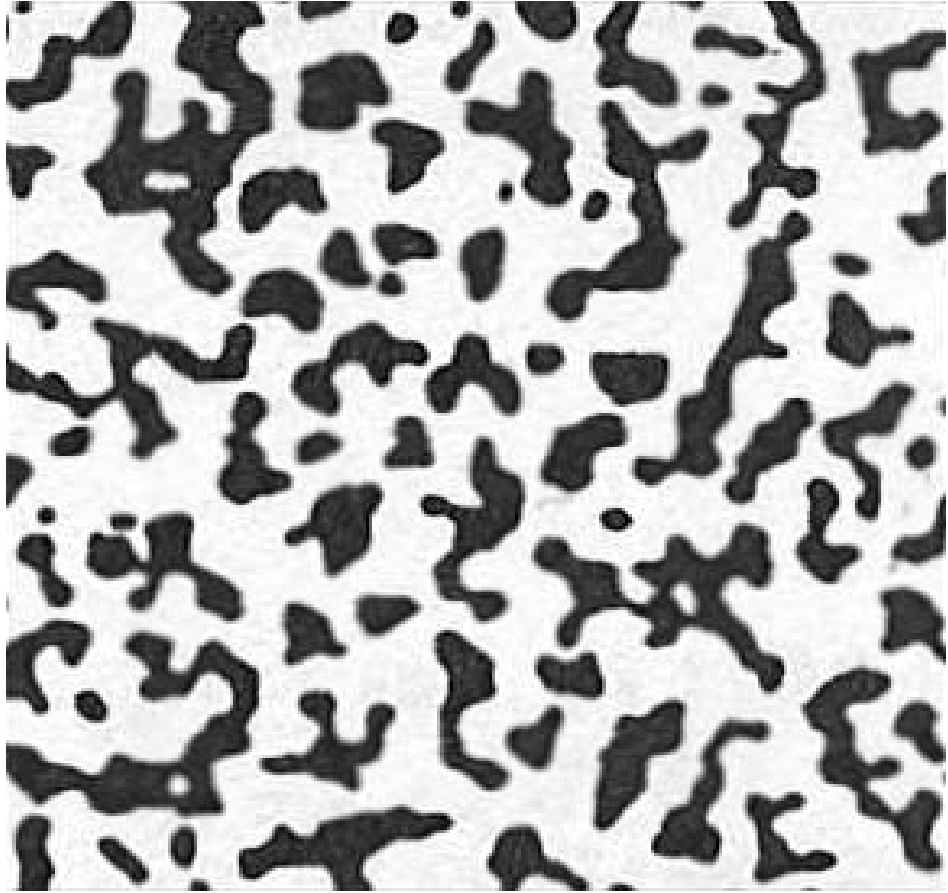

Thermoelectric Power of Bi and Bi-Te(0.14%) in Porous Vycor Glass  Huber *et al* .
Fig. 3

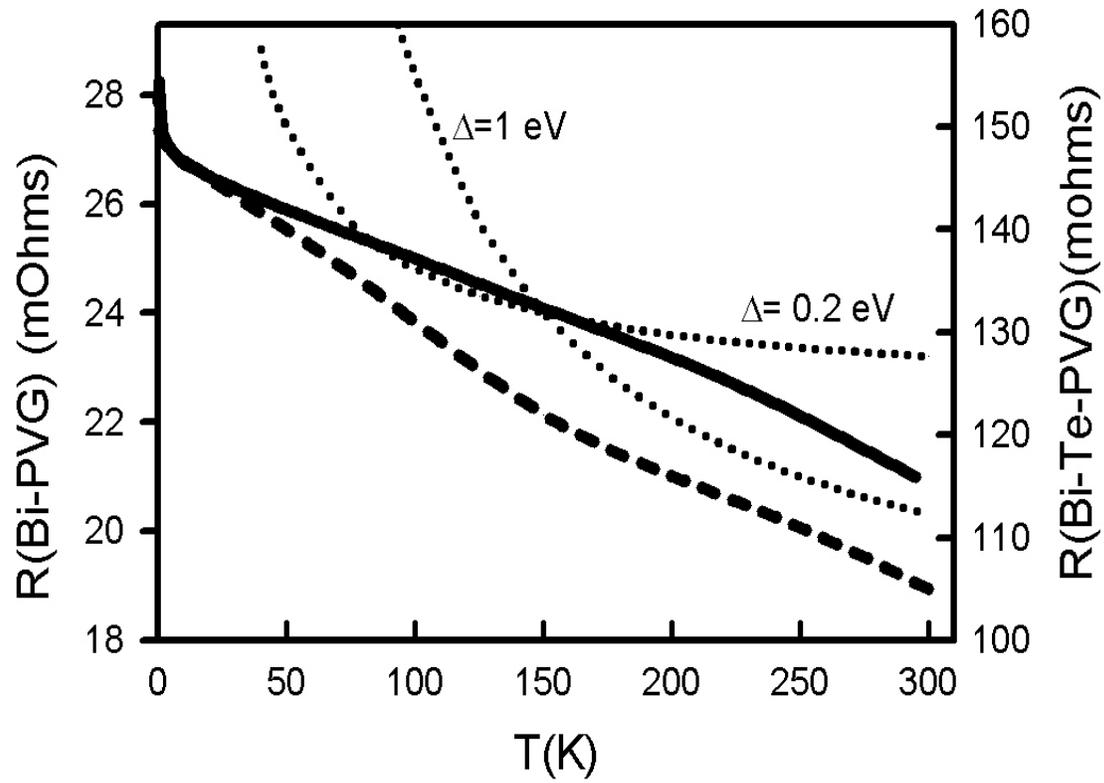

Thermoelectric Power of Bi and Bi-Te(0.14%) in Porous Vycor Glass  Huber *et al* .
Fig. 4

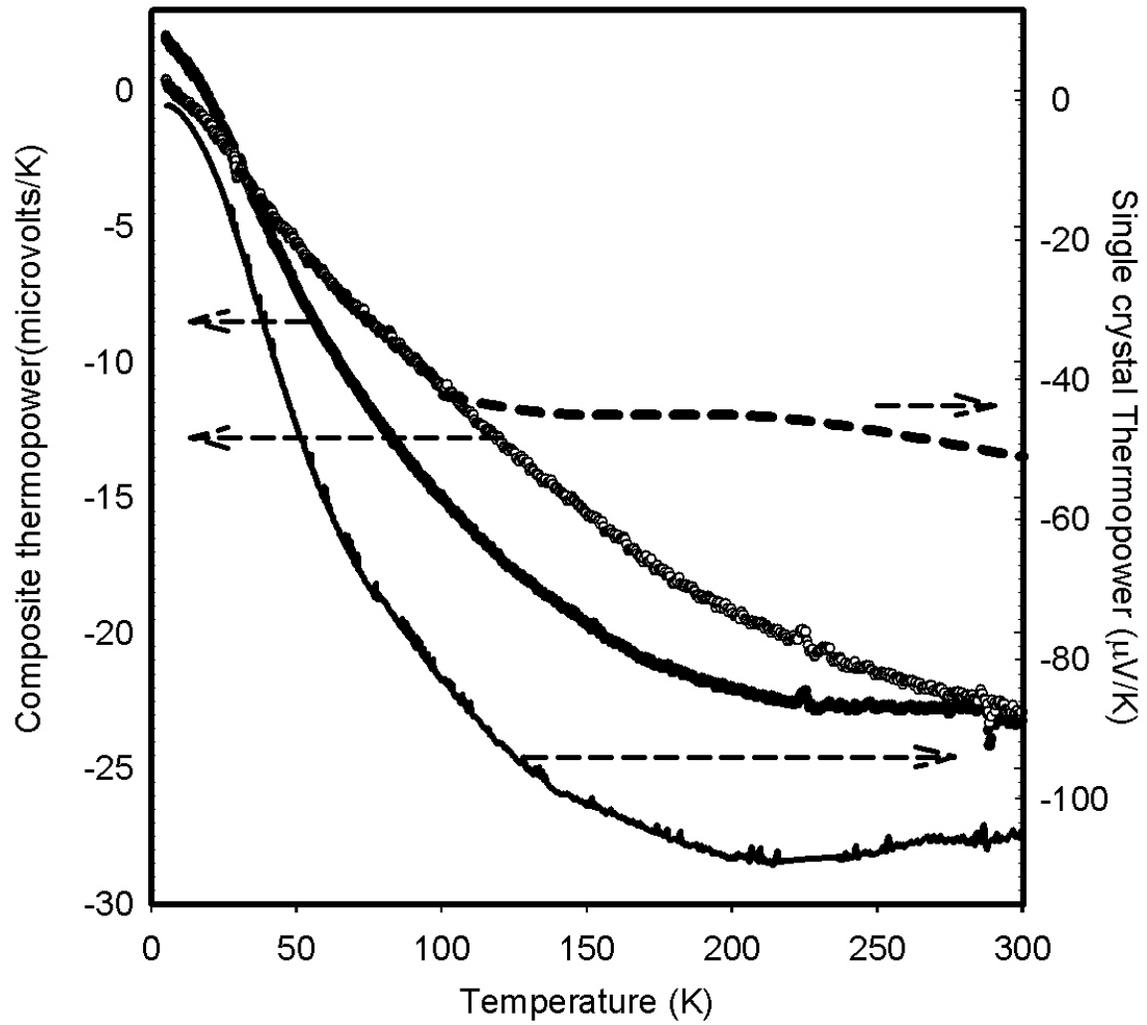

Thermoelectric Power of Bi and Bi-Te(0.14%) in Porous Vycor Glass  Huber *et al* .
Fig. 5